\documentclass[a4paper,12pt,epsf]{article}

\usepackage{t1enc}
\usepackage[latin1]{inputenc}
\usepackage[final]{graphicx}
\usepackage{amsfonts}
\usepackage{amssymb} 
\usepackage{oldgerm} 

\newcommand{\monu}{\mathfrak{u}}
\newcommand{\monw}{\mathfrak{w}}

\title{\LARGE \bf On certain perturbations of the Erd\"{o}s-Renyi
  random graph} 
\date{} \author{}

\begin{document}

\maketitle

\vspace{-1.2cm}

\centerline{\large St\'ephane Coulomb\footnote[1]{Email:
    coulomb@spht.saclay.cea.fr} and  Michel Bauer\footnote[2]{Email:
    bauer@spht.saclay.cea.fr}}

\vspace{.3cm}

\centerline{\large Service de Physique Th\'eorique de
  Saclay\footnote[3]{\it Laboratoire de la Direction des Sciences de
    la Mati\`ere du Commisariat \`a l'Energie Atomique, URA2306 du CNRS}}

\vspace{.3cm}

\centerline{\large CE Saclay, 91191 Gif sur Yvette, France}

\vspace{.3cm}

\begin{abstract} 
We study perturbations of the Erd\"{o}s-Renyi model for which the
statistical weight of a graph depends on the abundance of certain
geometrical patterns. Using the formal correspondance with an exactly
solvable effective model, we show the existence of a percolation
transition in the thermodynamical limit and derive perturbatively the
expression of the threshold. The free energy and the moments of the
degree distribution are also computed perturbatively in that limit and 
the percolation criterion is compared with the Molloy-Reed criterion.

\end{abstract}

\section{Introduction}

Random graphs were introduced more than forty years ago by
mathematicians and have proved since then to be a very useful and
versatile concept. The most studied
example is the Erd\"{o}s-Renyi model \cite{erdos},
where the edges are independent. Balanced with the simplicity of its
definition, the richness and deepness of mathematical results are really
fascinating.

On the other hand, it is clear that the Erd\"{o}s-Renyi model is a
poor idealization of real networks, those which pop out naturally in
sociology, biology, communication sciences,... For instance, the
degree distribution (i.e. the statistics of the number of edges
incident at a vertex) of most of the real life examples exhibits
statistical, scale-free, properties very far from the poissonian
behavior predicted by the Erd\"os-Renyi model \cite{barab},\cite{doro1}.

Many random graph models are now on the market, some consructed ad hoc
to reproduce certain desired features needed to fit real data, some
constructed according to general principles. Belonging to the second
category, the Molloy-Reed model \cite{molloy} concentrates, inside the space of all
labeled graphs with uniform probability, on the subspace of graphs
with an arbitrarily given degree distribution. Within this model, many
relevant quantities can be computed analytically, and there is a
general percolation criterion given in terms of cumulants of the edge
degree distribution.

Our aim is to study another family of random graphs for which explicit
computations are also possible. The idea is roughly as follows.
Suppose that to each graph $G$ one assigns a weight $u(G)$. From the
weight $u$ one can construct another weight $w(G)=\sum_{G' \subset G}
u(G)$ where the sum is over graphs $G'$ with the same vertex set as
$G$ and edge set included in that of $G$. Conversely, from any weight
function $w$ one can extract a unique weight function $u$, but the
expression of $u$ in terms of $w$ involves minus signs. 

We shall introduce a model for which the weight $u$ is a counting function for
certain structures on graphs. This weight $u$ has two further
properties : first it is permutation invariant, i.e. the weight of a graph
does not depend on the labelling of its vertices, and second it factors
over connected components, i.e the weight for a graph with several
connected components is the product of the weight of each component. 
Note that by standard combinatorial arguments, these three properties
are inherited by the weight $w$. 

Then we study the thermodynamic finite connectivity limit, when the
size of the system (i.e. the number of vertices of the graph) becomes
large but the average number of neighbors of any given vertex has a
fixed finite value. In this regime, the relevant features of the
weights $u$ and $w$ are encoded in tree generating functions $\monu$
and $\monw$ respectively.

The idea is that because $u$ counts
less objects than $w$ (which is not true for arbitrary $w$ because
then $u$ does not have a simple combinatorial interpretation in
general), the generating function $\monu$ has better convergence
properties than the corresponding generating function $\monw$. We
shall make the (crucial) assumption that the first
singularity in the generating function of $w$ can be obtained from
the functional relation that ties it to the generating function of
$u$, without having to know the singularities of the generating
function for $u$ itself. This is certainly true, as we shall recall
later, for the Erd\"os-Renyi model. It is also true order by order in
perturbation theory around the Erd\"os-Renyi model for the models we
introduce. We shall have little to say analytically on non
perturbative properties, but the numerical simulations are
encouraging.

Under this assumption we are able to give expressions for the
free energy, the size distribution of connected components and for the
percolation criterion and size of the giant component when it exists.
The expressions are not very explicit, because they involve the
function $\monu$, which is very complicated in terms of the original
parameters of the model. So on the one hand we show how to perform
explicit perturbative computations of the physical quantities and on
the other hand we introduce an effective model for which the
relationship between $u$ and $\monu$ is directly computable,

Our motivations are the following. First the models we study form a
natural and reasonnably manageable family of random graph models. Our
point is to emphasize the connection with quantum field theory. We do
not claim that the relation is very deep, but many random graph
phenomena have quantum field theory counterparts, and quantum field
theory  gives a very convenient language and insight. 
Second, one of the interests of studying models with non trivial
degree correlations is that attacks (see e.g. \cite{callaway},\cite{cohen})
automatically induce such features, even if they were absent to begin
with. Third, at a more basic level, we can contrast with the Molloy-Reed model. This is useful for
the purpose of general comparison, but especially 
because heuristic arguments, always based on non explicit assumptions,
allow to recover the Molloy-Reed percolation criterion whitout using
the particular hypothesis of the Molloy-Reed model, thereby suggesting
that the Molloy-Reed percolation criterion has a much wider range of
validity.  This is probably wrong, and the model solved in this paper
is definitely not in this range. 

\section{The model}

After recalling the elementary graph theoretic definitions, we present our
basic assumptions. We use the framework of statistical mechanics,
i.e. we assign to each labelled graph of size $N$ a weight (real
positive number), which we use as an unnormalized probability
distribution. For the Erd\"os-Renyi model, the weight is simply 
$p^{E(G)}(1-p)^{\frac{N(N-1)}{2}-E(G)}$ where $E(G)$ is the
number of edges of $G$. We shall choose a weight function that depends
on more detailed local features of the graph, namely the abundance of
certain geometric motives. 

\subsection{A few definitions and notations}

\paragraph{Simple unoriented graphs, connected graphs, trees.}
A (simple unoriented) graph $G$ is a couple $(V,E)$ where $V\neq
\emptyset$ is the vertex set and $E\subset \{\{i,j\}; i,j \in V, i
\neq j\}$ is the edge set. If $V=\{1,\cdots,N\}$ for some integer $N$,
then $G$ is called a labelled graph.  The set of labelled graphs of
size $N$ is denoted ${\mathcal G}_N$.
\\
If $G$ is a graph, we denote by $V(G)$ the vertex set of $G$ or the
cardinal of this vertex set, depending on the context, i.e whether a
set or a number is expected at that place \footnote{This should cause no
confusion, though from a fundamental point of view, a number is a
set as well.}. Similarly, $E(G)$ will denote either the edge set of $G$ or the
cardinal of this edge set.
\\
A connected component of $G$ is a minimal graph $(V',E')$ with $V'
\subset V$ such that if $(i,j) \in V'\times V$ and $\{i,j\}\in E$ then
$j \in V'$ and $\{i,j\}\in E'$.
\\
A connected graph is a graph which has only one connected component.
\\
A circuit of $G$ of size $s$ is a sequence $(i_1,\cdots,i_s)$ of
distinct vertices with $s \geq 3$ such that
$\{i_1,i_2\},\cdots,\{i_{s-1},i_s\}$ and $\{i_s,i_1\}$ are edges.
\\
A tree is a connected graph without circuits, and the set of labelled
trees of size $N$ is denoted by ${\mathcal T}_N$.
\\
If $\{i,j\}$ is an edge of $G$, we say that $i$ and $j \in V$ are
neighbours in $G$. The number of neighbours of a given vertex $i \in V$
in a graph $G$, also called the degree of $G$ at vertex $i$, is
denoted $l_i(G)$, or $l_i$ when there is no ambiguity.  It is the
number of elements of $E$ in which $i$ appears.

\paragraph{Adjacency matrix of a graph, operations on matrices.}

The adjacency matrix $A(G)$ (or simply $A$) of a labelled graph 
$G \in {\mathcal G}_N$ is the $N$ by $N$ matrix defined by $A_{i,j}=1$ if
$\{i,j\}$ is an edge of $G$, $A_{i,j}=0$ else. Note that the set of
adjacency matrices is the set of symmetric $0,1$ matrices with
vanishing diagonal elements.
\\
The sum of all elements of any matrix $M$ will be written 
$\| M \| \equiv \sum_{i,j} M_{i,j}$. If $M$ is the adjacency matrix 
$A(G)$ of a graph $G$, then it is clear that $\frac{1}{2} \| M \|$ 
is $E(G)$, the number of edges of $G$.
\\
The sum of all diagonal elements of a square matrix $M$ is the trace
of $M$, written $Tr(M)$. Note that $\frac{1}{2} Tr(A(G)^2)$ is again
equal to $E(G)$.

\subsection{Probability distribution, partition function}

\label{probdist}

To emphasize the similarities between the random graph model
studied in this paper and quantum field theory, we split the weight of
graphs in a product of a free part and an interacting part. The free
weight is $w_0(G)\equiv q^{E(G)}$ where $q \in ]0,+\infty[$. For later
convenience, we also introduce $p=q/(1+q) \in ]0,1[$. The interacting
part is $w_I(G)\equiv e^{S_I(G)}$, where
$$S_I(G)\equiv \sum_k \frac{t_k }{2k}Tr A(G)^k+\sum_k \frac{s_k}{2}
\|A(G)^k\|.$$
The full weight is $w(G)\equiv w_0(G) w_I(G).$ The
normalization factor included in the definition of the partition
function
$$Z_N \equiv (1-p)^{\frac{N(N-1)}{2}}\sum_{G \in {\mathcal G}_N}
w(G)$$
is chosen in such a way that $Z_N$ can be expressed as an
average over Erd\"o{s}-Renyi weights :
$$Z_N = \left< e^{S_I(G)}
\right>_{ER(p)},$$
where $ER(p)$ assigns probability $p^E (1-p)^{\frac{N(N-1)}{2}-E}$ to
any graph on $N$ vertices with $E$ edges.
We view $w_0$ as describing a gas of independent edges (the
Erd\"o{s}-Renyi model), and $S_I$ as describing the interactions
between edges, the $t_k$'s and $s_k$'s being arbitrary real parameters
that regulate the abundance of certain local geometric features of
$G$. Note that if $G$ is made of several connected components,
$G_1,\cdots,G_l$, $S_I(G)=S_I(G_1)+\cdots+S_I(G_l)$, so that $w_I$
factors as a product over connected components. This is also true of
$w_0$. This multiplicativity of the weight plays a crucial role to
simplify the analysis below. Another striking feature is that $w(G)$
is invariant under permutations of the vertex set. Many other
interactions with this property could be incorporated, for instance by
including products of traces and norms into the interaction, however
these would break the multiplicativity property. The ``simple''
multiplicative model does not seem to be exactly solvable, and we have
to rely on perturbation theory to make explicit computations. Even the
``trival'' case when all parameters except, say, $s_3$ vanish is
already complicated enough. This is why we insist on keeping the two
properties : multiplicativity and permutation invariance.

The above model is perfectly well defined for all parameter values as long
as $N$ is finite. However, we shall be interested in taking a large
$N$ limit such that the average degree is a fixed number (so that the
number of edges is proportional to $N$). This will impose some
constraints, see below.

\section{Two combinatorial formul\ae\ for the partition function}

In this section, we shall derive two formul\ae\ related to
exponentiation in combinatorics.

\subsection{Connected components}

Suppose we consider the grand canonical partition function
$$\Xi=\sum_G \frac{w(G)}{V(G)!}z^{V(G)}$$
as a power series in $z$ :
this is a formal sum over all graphs of any size, from which $Z_N$
can be recovered as
$$Z_N=(1-p)^{\frac{N(N-1)}{2}} N! \oint \frac{dz}{z^{N+1}}\Xi(z)$$
where for the time being, the symbol $\oint \frac{dz}{z^{N+1}}$ is not
viewed as a real contour integral, but simply as the operation of
taking the term of order $N$ in the $z$ expansion of a formal power
series. 
If $G$ has $l$ connected components $G_1,\cdots,G_l$, we observe that

$$\frac{w(G)}{V(G)!}z^{V(G)}=\frac{1}{\left(\sum_{k=1}^l
    V(G_k)\right)!}  \prod_{k=1}^l q^{E(G_k)}w_I(G_k)z^{V(G_k)}.$$
Summing over $G$ is the same as summing over the connected components.
Up to now the vertices of the $G_k$ were labelled as subgraphs of $G$.
Thanks to permutation invariance, one can instead sum over abstract
finite sequences of $l=1,2,\cdots$ labelled connected graphs but
weight the terms in the sum by a combinatorial factor
$$\frac{1}{l!}\frac{\left(\sum_{k=1}^l V(G_k)\right)!}{\prod_k
  V(G_k)!}$$
to take into account all possibilities to order them and
to label the union of their vertex sets from $1$ to $\sum_k V(G_k)$.
Then $\Xi=e^W$ where $W= \sum^c_G \frac{w(G)}{V(G)!}z^{V(G)}\equiv
\sum_n \frac{W_n}{n!}z^n$ is defined exactly as $\Xi$ except that the
sum is only over connected graphs (this is what is meant by the symbol
$\sum^c_G$). The main formula of this section, the first exponential
formula, relates $Z_N$ and $W(z)$ as
\begin{equation}
  \label{eq:1expform}
Z_N=(1-p)^{\frac{N(N-1)} {2}} N! \oint \frac{dz}{z^{N+1}}e^{W(z)}.  
\end{equation}

\subsection{Reorganization of the perturbative expansion}
\label{reorg}

The case when $t_k=s_k=0$ for all $k$ corresponds to the
Erd\"{o}s-Renyi model. The probability of a graph only depends on the
number of its edges, and many quantities such as degree distributions,
component distributions and percolation threshold take a simple form.
When one or more of the $t_k$'s and $s_k$'s are not vanishing, the
finer structure of the graph becomes relevant, and this will be the
case of interest in this paper.

Let us fix $G$ and start from the expansion of $e^{S_I(G)}$ in powers
of $t_k$'s and $s_k$'s. This gives a linear combination of terms of the
form:
$$\prod_k \left( \left( Tr A(G)^k \right)^{m_k} \| A(G)^k \|^{n_k}
\right).$$
If we expand each matrix product, such a term becomes a sum
of products of matrix elements of $A(G)$ of generic form $
A(G)_{i_1j_1} A(G)_{i_2j_2} \cdots A(G)_{i_nj_n} $ and we may assume
that $i_1\neq j_1,\cdots,i_n\neq j_n$ because otherwise the product is
$0$ for any (simple graph) adjacency matrix.  On the other hand, to
any sequence $i_1j_1 \cdots i_nj_n$ with $i_1\neq j_1,\cdots,i_n\neq
j_n$, we may associate a graph $H$ with vertex set $[1,N]$ and edge
set $\left\{\{i_1,j_1\},\cdots,\{i_n,j_n\}\right\}$. The product
$A(G)_{i_1j_1} A(G)_{i_2j_2} \cdots A(G)_{i_nj_n}$ vanishes unless all
edges of $H$ are edges of $G$, in which case it has value 1.

With this observation in mind, we define $\overline{e^{S_I(G)}}$ by
keeping, in the expansion of $e^{S_I(G)}$, only those terms $
A(G)_{i_1j_1} A(G)_{i_2j_2} \cdots A(G)_{i_nj_n}$ such that
$\left\{\{i_1,j_1\},\cdots,\{i_n,j_n\}\right\}$ exhausts the edge set of
$G$ (maybe with repetitions). Then by definition, 
$$e^{S_I(G)}=\sum_{H, E(H) \subset E(G)} \overline{e^{S_I(H)}},$$
where the sum is over all graphs on the same vertex set as $G$ whose
edge set is a subset of that of $G$. The reciprocal formula is given
by $ \overline{e^{S_I(H)}}=\sum_{G, E(G) \subset E(H)}(-)^{E(H)-E(G)}
e^{S_I(G)}$ and the multiplicative property of $e^{S_I(G)}$ ensures
that $\overline{e^{S_I(H)}}$ is also multiplicative : if $H$ has $l$
connected components $H_1,\cdots,H_l$,
$$\overline{e^{S_I(H)}}=\overline{e^{S_I(H_1)}}\cdots
\overline{e^{S_I(H_l)}}.$$

Now
\begin{eqnarray*}
Z_N & = & \sum_{G \in {\mathcal G}_N}
(1-p)^{\frac{N(N-1)} {2}-E(G)}p^{E(G)}e^{S_I(G)} \\
& = & \sum_{G \in {\mathcal G}_N} \sum_{H, E(H) \subset E(G)}
(1-p)^{\frac{N(N-1)} {2}-E(G)}p^{E(G)} \overline{e^{S_I(H)}}\\
& = & \sum_{H \in {\mathcal G}_N} \sum_{G, E(G) \supset
  E(H)}(1-p)^{\frac{N(N-1)} {2}-E(G)}p^{E(G)} \overline{e^{S_I(H)}}. 
\end{eqnarray*}
For fixed $H$, $\sum_{G, E(G) \supset E(H)} (1-p)^{\frac{N(N-1)} 
 {2}-E(G)}p^{E(G)}=p^{E(H)}$, and we find that 
$$Z_N=\sum_{H \in{\mathcal G}_N} p^{E(H)}\overline{e^{S_I(H)}}.$$
Defining $$u(H)=p^{E(H)}\overline{e^{S_I(H)}} \qquad U=\textstyle{\sum^c_H}
\frac{u(H)}{V(H)!}z^{V(H)}\equiv \sum_n \frac{U_n}{n!}z^n,$$ we can
repeat the steps leading from multiplicativity and permutation
invariance to eq.(\ref{eq:1expform}) to obtain the second exponential
formula
\begin{equation}
  \label{eq:2expform}
Z_N= N! \oint \frac{dz}{z^{N+1}}e^{U(z)}.    
\end{equation}

\subsection{Consequences of the exponential formul\ae}

The two expressions obtained for the partition function, one in terms
of $W(z)$ and the other in terms of $U(z)$, show that for every
nonnegative integer $N$
$$ (1-p)^{-\frac{N(N-1)} {2}} \oint \frac{dz }{z^{N+1}} e^{U(z)} = \oint
\frac{dz } {z^{N+1}} e^{W(z)}.$$

It is convenient to eliminate $N$  by the following trick :
putting $1-p=e^{-\beta}$, $(1-p)^{-\frac{N(N-1) } {2}}$ takes the form
of a gaussian integral $\frac{1} {\sqrt{2\pi \beta}}
\int_{-\infty}^{+\infty} e^{-\frac{y^2 } {2\beta}+y 
N-\beta \frac{N} {2}}$. From the change of variable $x=z
e^{y-\frac{\beta } {2}}$, it follows that
\begin{equation}
e^{W(z)}=\frac{1} {\sqrt{2 \pi \beta}}  \int_{-\infty}^{+\infty}
e^{-\frac{y^2 } {2\beta}+U(z e^{y-\frac{\beta } {2}})}dy.
\label{eq:egalexact}
\end{equation}
There is no good reason why $e^{-\frac{y^2 } {2\beta}+
U(z e^{y-\frac{\beta } {2}})}$ should be integrable in $y$ along
the full real axis. However, if one expands this function in powers
of $z$, term by term integration is ok, and for the time being,
eq.(\ref{eq:egalexact}) is a shorthand notation for the fact that this
term by term integration leads to the formal power series of
$e^{W(z)}$.

\section{Practical perturbative expansion}

Our aim is to organize the perturbative expansion to make explicit
computations. We would like to make a systematic enumeration of the
terms that appear in perturbation theory. 

A typical term in the perturbative expansion is of the form $A_{i_1j_1}
 \cdots A_{i_nj_n}$ to which we associate the sequence $i_1j_1 \cdots
i_nj_n$ i.e. a word written using the alphabet $[1,N]$. For a graph $G$ with
adjacency matrix $A_{ij}$, the product $A_{i_1j_1}
A_{i_2j_2} \cdots A_{i_nj_n}$ is $1$ if
$\{i_1,j_1\},\cdots,\{i_n,j_n\}$ are amongst the edges of $G$ and $0$
else. If $l$ is the number of distinct edges among these $n$ 2-sets,
the average is simply the sum of the Erd\"{o}s-Renyi weights of all
graphs containing these $l$ edges. This is known to yield $\left<
  A_{i_1j_1} A_{i_2j_2} \cdots A_{i_nj_n} \right>_{ER(p)}=p^l$.  This
average is invariant under permutations of $[1,N]$, 
all vertices play the same role in $Tr A(G)^k$ and $\| A(G)^k
\|$. So we regroup the words $i_1j_1 \cdots
i_nj_n$ in classes under the action of the permutation group, compute
the size of each class and find a representative in each class. Then
we enumerate the representatives and take multiplicities into account.

The idea is the following : suppose that you have a finite word
written using any alphabet (i.e. any set of symbols) on, say, N
letters. To each letter that appears in the word, associate an
integer as follows : assign $1$ to the first letter of the word, then
assign $2$ to the next new (i.e distinct from the first) letter
appearing in the word, then $3$ to the next new (i.e distinct from the
first and the second) and so on until all letters appearing in the
word have been assigned a number, the highest one being, say, $v$ ($v$
is the number of distinct letters used to compose the word, which may
well be strictly smaller than the length of the word, because the
same letter can appear more than once). Replacing each letter of the
word by its number leads to a new word, the alphabet being $[1,v]
\subset [1,N]$ this time. The words obtained by this procedure are
characterized by the fact that $1$ appears before $2$ which appears
before $3$ and so on. Say that two words in the original alphabet are
equivalent if they yield the same numerical word by the above
procedure. Then each class contains $\frac{N!}{(N-v)!}$ words. 

In our case, the original alphabet is already $[1,N]$, and we are led
to the concept of normalized sequences, an elaboration of a procedure
introduced in a slightly simpler context in \cite{bg}.

\subsection{Normalized sequences} 

For an arbitrary sequence $i_1j_1 i_2j_2\cdots i_nj_n$ (with alphabet
$[1,N]$) such that $i_1\neq j_1,\cdots,i_n\neq j_n$, we define
$v=\#\{i_1,j_1,\cdots,i_n,j_n\}$, the number of distinct vertices in
the sequence, and $l=\#\{\{i_1,j_1\},\{i_2,j_2\},\cdots,\{i_n,j_n\}\}$,the number
of distinct edges in the sequence.
  
We shall say that a sequence $i_1j_1 i_2j_2\cdots i_nj_n$ is
normalized with respect to $\prod_k
  \left( Tr A(G)^k \right)^{m_k} \prod_k \| A(G)^k \|^{n_k}$ or more
  simply with respect to  $(m_k,n_k)$ if
\begin{itemize}
\item $n=\sum_k k(m_k+n_k)$.
\item In this sequence, $1$ comes before $2$, which comes before
  $3$,... which comes before $v$.
\item $i_1 \neq j_1,\cdots,i_n \neq j_n$.
\item The sequence has a correct structure as regards $Tr$ and $\| \ 
  \|$. That is, to $Tr A(G)^p$ ($m_k=\delta_{p,k},n_k=0$) and
  $\|A(G)^p\|$ ($m_k=0,n_k=\delta_{p,k}$) correspond the constraints
  $j_1=i_2,\cdots,j_{p-1}=i_p$, with the additional constraint
  $j_p=i_1$ for $Tr A(G)^p$. If more than one term of $(m_k,n_k)$ is nonzero,
  then we choose an arbitrary ordering : increasing $k$'s, all traces
  coming before norms. This allows to decompose the sequence in
  subsequences, which correspond either to a trace or a norm, and are
  accordingly constrained. For instance, when $m_3=1$ and
  $n_1=1,n_3=1$ are the only nonvanishing elements of $(m_k,n_k)$,
  the sequence has a correct structure if it is of the form
  $i_1j_1i_2j_2i_3j_3 \ i_4j_4 \ i_5j_5i_6j_6i_7j_7$ where
  $j_1=i_2,j_2=i_3,j_3=i_1$ and $j_5=i_6,j_6=i_7$ (to $\|A(G)\|$
  correspond no constraint of structure).
\end{itemize}

We write ${\mathcal M}_{v,l,(m_k,n_k)}$ for the number of normalized
sequences with $v$ vertices and $l$ edges. By our previous remarks,
the class containing a normalized sequence has $\frac{N!}{(N-v)!}$
members, each of which leads to the same average. Hence

\begin{equation} \label{eq:avnorm} \left< \prod_k \left( \left( Tr
A(G)^k \right)^{m_k} \| A(G)^k \|^{n_k} \right) \right>_{ER(p)}=
\sum_{v,l}\frac{N!} {(N-v)!} p^l {\mathcal M}_{v,l,(m_k,n_k)}.
\end{equation}

In doing explicit computations, which can be painful, there is a
useful check of the formula, namely a sum rule corresponding to $p=1$,
in which case only the complete graph contributes, and there is no
average to compute. It is staightforward to check that if $G$ is the
complete graph on $N$ vertices,
$$Tr A(G)^k= (N-1)^k+(-)^k(N-1) \qquad \| A(G)^k \| =N(N-1)^k.$$
Hence 
$$\sum_{v,l}\frac{N!} {(N-v)!}{\mathcal M}_{v,l,(m_k,n_k)}=\prod_k
\left((N-1)^k+(-)^k(N-1)\right)^{m_k}\left(N(N-1)^k\right)^{n_k}.$$

\subsection{Graphical expansion}

\begin{figure}
  \begin{center}
    \includegraphics[width=.95\textwidth]{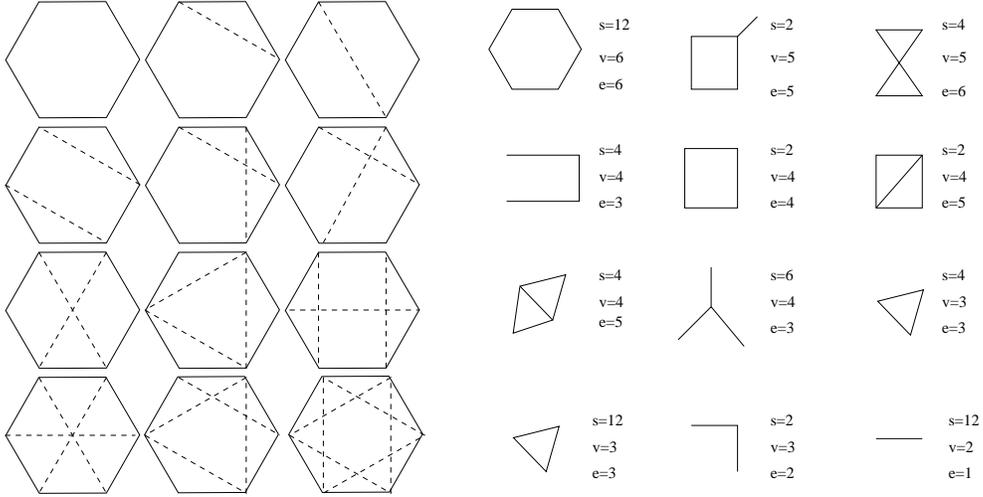}
    \caption{Graphical enumeration/interpretation of the contribution
      of $Tr A^6$ to the partition function. On the left are drawn all
      possible identifications, schematized by complete dashed graphs.
      On the right identification has been carried out. The numbers of
      symmetries, vertices and edges are given for each graph.}
    \label{coefft6}
  \end{center}
\end{figure}

Although the interpretation in terms of normalized sequences is
adequate for the purpose of numerical computations, there is another
useful graphical representation of the perturbation series which we
present briefly now. \\
Expanding $Tr A^k=\sum_{i_1,\cdots,i_k} A_{i_1,i_2}A_{i_2,i_3}\cdots
A_{i_k,i_1}$ we represent each term as a colouring of a labelled cycle
on $k$ vertices with $N$ colours, vertex $j$ carrying color $i_j$ for
$j=1,\cdots,k$. In the same way, we represent each term in
$\|A^k\|=\sum_{i_1,\cdots,i_{k+1}} A_{i_1,i_2}A_{i_2,i_3}\cdots
A_{i_k,i_{k+1}}$ as a colouring of a labelled segment on $k+1$
vertices with $N$ colours, vertex $j$ carrying color $i_j$ for
$j=1,\cdots,k+1$. The expansion of $e^{S_I}$ in powers of $t_k$'s and
$s_k$'s, is then represented as a sum over colorings, with $N$
colours, of labelled graphs whose connected components are cycles and
segments. Pick one term, call it $\Gamma$, in this sum. Each cycle of
length $k$ yields a factor $t_k/(2k)$, each segment on $k+1$ vertices
yields a factor $s_k/2$, there is a factor $1/m_k!$ if there are $m_k$
cycles of length $k$, and a factor $1/n_k!$ if
there are $n_k$ segments on $k+1$ vertices. \\
The probabilistic average of $\Gamma$ (over the set of incidence
matrices) is zero if some edge of $\Gamma$ has it's two extremities of
the same colour. If not, let $e(\Gamma)$ be the number of distinct
pairs of colors that appear as extremities of edges of $\Gamma$. The
probabilistic average over the set of incidence matrices multiplies
the former weight of $\Gamma$ by $p^{e(\Gamma)}$. \\
Let $v(\Gamma)$ be the number of distinct colours in the colouring of
$\Gamma$. Say that terms $\Gamma$ and $\Gamma'$ are equivalent if
there is a permutation of $[1,N]$ (the set of colours) that maps
$\Gamma$ to $\Gamma'$. The equivalence class of $\Gamma$ is made of
$N!/(N-v(\Gamma))!$ graphs with the same weight. The equivalence class
of $\Gamma$ has a graphical representation : starting from $\Gamma$,
draw a dashed line between two vertices if and only if they carry the
same color. Then remove the colors. In this way, obtain a graph with
two kind of edges, solid and dashed. The graphs that appear in this
operation have two properties. First, the solid components are cycles
and segments, and the dashed components are complete graphs. Second,
two vertices cannot be adjacent for solid and dashed edges at the same
time. In lack of a better denomination, we call graphs satisfying
these two conditions (labelled) admissible graphs. The notion of
connectivity for admissible graphs treats solid and dashed edges on
the same footing. \\
Instead of working with labelled admissible graphs, we may use
unlabelled admissible graphs. Then the combinatorial factors ($2k$ for
a $k$-cycle, $2$ for a $k+1$-segment, a factorial for permutations of
components of the same type and size) which take into account only
solid edges, are replaced by the order of the symmetry group of the
admissible graph, the group
of permutations of vertices that preserve solid and dashed edges.\\
Let $H$ be an admissible graph. Two vertices being declared equivalent
if they are connected by a dashed line, let $v(H)$ be the number of
equivalence classes of vertices. Two edges being declared equivalent
if their extremities are equivalent as vertices, let $e(H)$ be the number
of equivalence classes of edges. Furthermore, we denote by $s(H)$ the
order of the symmetry group of the graph. Then
$$Z_N=\sum _H \frac{1}{s(H)}p^{e(H)} \frac{N!}{(N-v(H))!}\prod _k
t_k^{m_k(H)}s_k^{n_k(H)},$$
where the sum is over unlabelled
admissible graphs $H$, $m_k(H)$ is the number of solid $k$-cycles and
$n_k(H)$ the number of solid segments on $k+1$ vertices.
See figure \ref{coefft6} for the example of $Tr A^6$.

\subsection{Exponentiation}

\label{sec:expon}

Computations on a sheet of paper are more economical using admissible
graphs, but systematic machine enumeration is best carried out using 
normalized sequences.
\\
If is not difficult to convince oneself that the notion of
connectedness of normalized sequences or of admissible graphs is the
same, and coincides with the notion of connectedness used to establish
eq.(\ref{eq:2expform}). We infer that
$$U(z)=z+\textstyle{\sum^c_H}
\frac{1}{s(H)}p^{e(H)} z^{v(H)} \prod _k t_k^{m_k(H)}s_k^{n_k(H)},$$
where $\sum^c_H$ is the sum over unlabelled connected admissible
graphs or equivalently that
$$U(z)=z+\sum_{v,l,\{m_\bullet,n_\bullet \}} z^v p^l
\frac{\tilde{t}_\bullet^{m_\bullet}  \tilde{s}_\bullet^{n_\bullet} }
 {m_\bullet ! n_\bullet !}
   {\mathcal M}_{v,l,\{m_\bullet,n_\bullet \}}^c,$$ 
where $\tilde{t}_\bullet^{m_\bullet} \tilde{s}_\bullet^{n_\bullet}
\equiv \prod_k \left( \frac{t_k } {2k } \right)^{m_k} \left(
  \frac{s_k} {2} \right)^{n_k}$ and ${\mathcal M}_{v,l,(m_k,n_k)}^c$
is the number of normalized connected sequences.

\section{Finite connectivity large $N$ behaviour}

Our aim is to use the identities
eqs.(\ref{eq:1expform},\ref{eq:2expform}) and their consequence
eq.(\ref{eq:egalexact}) to derive mean field type identities valid in
the limit $N \rightarrow \infty$, $pN,t_k,s_k$ being $N$ independant,
or more generally having finite limits for large $N$. It is customary
to define $\alpha \equiv \lim_{N \rightarrow \infty} pN$.

\subsection{Caveat}

In this regime, when $t_k=s_k=0$ for all $k$'s (the Erd\"os-Renyi
model), the event that a graph contains a component with much more
edges than vertices has a vanishingly small probability, and the
connected components look locally like trees. This is called the
dilute regime.  In particular, the complete graph (a caricature of a
non dilute graph) has a negligible weight in the dilute regime.

We are
looking for an analogous regime for the perturbed Erd\"os-Renyi model.
However, in that case, the complete graph has weight 
$$ w=q^{N(N-1)/2}e^{\sum_k \frac{t_k }{2k}((N-1)^k+(-)^k(N-1))+
  \sum_k\frac{s_k}{2} N(N-1)^k}.$$
Compare this to a
union of isolated vertices (a caricature of a dilute graph), which 
has weight $w=1$. Consider for instance, the case when there is only
one nonvanishing perturbation parameter, say $s_3$. The two weights
are equal if $qe^{s_3(N-1)}=1$. That this crude balance
gives the correct qualitative frontier between a dilute regime and a
dense regime is confirmed by numerical simulations.

If $p\equiv \alpha/N$ and
$s_3<0$, the complete graph is indeed strongly suppressed (in fact
much more drastically than for the pure Erd\"os-Renyi model). However,
if $s_3>0$, the weight of the complete graph submerges the weight of
dilute configurations. 

To summarize, the following discussion makes
sense only if $S_I$ does not become positive and large ($\gg N^2$)
for non dilute configurations. An easy way to ensure that is to take
the sign of all perturbations negative. Another possibility would be
to take only a finite number of nonzero perturbations, and then impose
that the dominant one be negative. 

\subsection{Main equations}
 \label{maineq}
With these observations in mind, we start from
\begin{eqnarray*}
Z_N  & = & N! \oint \frac{dy}{y^{N+1}}e^{U(y)}\\
& = & (1-p)^{\frac{N(N-1)} {2}} N! \oint \frac{dx}{x^{N+1}}e^{W(x)}
\end{eqnarray*}
and 
$$e^{W(x)}=\frac{1} {\sqrt{2 \pi \beta}} \int_{-\infty}^{+\infty}
e^{-\frac{z^2 } {2\beta}+U(x e^{z-\frac{\beta } {2}})}dz.$$
Recall
that $p=q/(1+q)=1-e^{-\beta}$. We fix $p=p_N\equiv \alpha/N$ where
$\alpha$ is a constant, and make changes of variables $y\rightarrow
y/p$, $x\rightarrow x/q$ in the above integrals. In the expansions
$$U(y/p)= \textstyle{\sum^c_H} 
\left(\frac{\alpha}{N}\right)^{E(H)-V(H)} 
\overline{e^{S_I(H)}}\frac{y^{V(H)}}{V(H)!}$$
and 
$$W(x/q)= \textstyle{\sum^c_H}
\left(\frac{\alpha}{N-\alpha}\right)^{E(H)-V(H)} 
e^{S_I(H)}\frac{x^{V(H)}}{V(H)!}$$
the sum is over connected graphs, and by Euler formula, $E-V=L-1$
where $L\geq 0$ is the number of loops. 
Hence we may write formaly
$$U(y/p)=\frac{N}{\alpha}\monu(y)+o(N) \qquad
W(x/q)=\frac{N}{\alpha}\monw(x)+o(N),$$ 
where $\monu(y)=\sum_{T} \overline{e^{S_I(T)}}\frac{y^{V(T)}}{V(T)!}$ and
$\monw(x)=\sum_{T} e^{S_I(T)}\frac{x^{V(T)}}{V(T)!}$ are sums over
trees (connected graphs with $L=0$).
If we use a na\"{\i}ve version of the saddle point approximation and
write $Z_N=e^{NF+o(N)}$, we find 
\begin{eqnarray}
\label{eq:mainu}
F & = & -1-\log
\frac{y^*}{\alpha}+\frac{1}{\alpha}\monu(y^*)\\ 
\label{eq:mainw}F & = & -\frac{\alpha}{2} -1-\log
\frac{x^*}{\alpha}+\frac{1}{\alpha}\monw(x^*)\\ 
\label{eq:mainuw} \monw(x) & = &-\frac{{\hat{z}}^2}{2}+\monu(xe^{\hat{z}}),  
\end{eqnarray}
where the $x^*,y^*$ and $\hat{z}$ are appropriate saddle point values :
$$\alpha=x^*\monw'(x^*)=y^*\monu'(y^*) \qquad
\hat{z}=xe^{\hat{z}}\monu'(xe^{\hat{z}}).$$

We end this section with the following remarks.  The average number of
edges is more or less the variable conjugate to $p$. More precisely,
the average number of edges is
$$
q \frac{\partial}{\partial q} \log \left((1-p)^{-N(N-1)/2}Z_N \right).$$
We infer that in the
thermodynamic regime with $N \rightarrow \infty$ and $p_N = \alpha/N$,
the average number of neighbors of a given point (i.e. 2/N times the
average number of edges) is
\begin{equation}
c = \alpha +2\alpha \frac{\partial F}{\partial \alpha} 
\label{cavecF}
\end{equation}
For the pure Erd\"os-Renyi model, the weights form a probability
distribution, $Z_N=1$, $F=0$, and $c=\alpha$. In the perturbed models,
$\alpha$ is not so easily measured on the graph, and only the
parameter $c$ has direct physical meaning. From the point of view of
quantum field theory, it is natural to view $\alpha$ as the bare
connectivity and $c$ as the physical connectivity. For each coupling
constant $t_k$ or $s_k$, it would be desirable to find analogous
physical quantities that first, one can compute directly on a random
graph without knowing a priori the sampling measure and that second one
can reduce $t_k$ or $s_k$ to first order in perturbation theory.  This
is very ambiguous and we have not found an elegant way to select such
physical observables systematically.

\subsection{Discussion}

We have seen before that a dilute regime for the perturbed
Erd\"os-Renyi model with fixed values of the $t_k$'s and $s_k$'s
cannot exist if $S_I$ becomes large positive for graphs with many
loops. Here we discuss a related limitation even if one considers
only loopless graphs. 

Instead of considering the complete graph, look
at the star shaped tree on $n$ vertices, whose adjacency matrix we
denote by $S$, with a center connected to the $n-1$
other vertices. From $Tr S =0$, $\| S\|= Tr S^2=2(n-1)$, $\| S^2 \|=n(n-1)$
and $S^3=(n-1)S$, it is easy to compute recursively that $Tr
S^{2k+1}=0$ and $\| S^{2k+1} \|=2(n-1)^{k+1}$
for $k \geq 0$,  and that $Tr S^{2k}=2(n-1)^k$ and $\| S^{2k}
\|=n(n-1)^k$ for $k \geq 1$.  
As an example, consider again the case when there is only
one nonvanishing perturbation parameter, say $s_3$. The contribution
of star shaped trees to $\monw$ is $\sum_n
\frac{1}{(n-1)!}e^{s_3(n-1)^2}x^n$.
As all trees give a positive contribution to $w$, no compensation is
possible and we conclude that if $s_3 >0$, the series for $w$ has a
vanishing radius of convergence. So it is meaningless to deform
contours, and eq.(\ref{eq:mainw}) is meaningless as well. Then so is
eq.(\ref{eq:mainu}) because analyticity of $\monu(y)$ at small $y$
implies analyticity of $\monw(x)$ at small $x$ via eq.(\ref{eq:mainuw}).
On the other hand, if $s_3 <0$, the star-shaped trees of large size are very
strongly suppressed. Let us note however as shown in the next section that, in the realm
of formal power series, eq.(\ref{eq:mainuw}) describes the correct
combinatorial relationship between $\monu$ and $\monw$ even if both series
have a vanishing radius of convergence.

More generally, if $S_I(T)/V(T)$ is bounded above (an easy way to
ensure that is to take the sign of all perturbations negative, another
possibility would be to take only a finite number of nonzero
perturbations, and then impose that the dominant one be negative) ,
$\monw$ is analytic near the origin. Indeed, if $S_I(T)/V(T) \leq \tau$
for all trees, using the fact that there are $n^{n-2}$ labelled trees
on  $n$ vertices, we see that $0 < \frac{w_n}{n!} \leq
\frac{n^{n-2}}{n!}e^{\tau n}$, leading to a nonzero radius of
convergence. 

For instance, when the sign of every perturbation is negative, the
radius of convergence is a nonincreasing function of the $t_k$'s and
$s_k$'s : it gets larger and larger as the $t_k$'s and $s_k$'s get
more negative. To see that it remains finite,
consider the linear graph on $n$ vertices, whose adjacency matrix we
denote by $L$. For this graph, for fixed $k$ and large $n$, $Tr L^k $
and $\| L^k \|$ grow at most linearly with $n$ : they count $k$ steps
walks, and if the starting point is given, at each step there are at
most two choices, so there is the obvious upper bound $n2^k$. There are
$n!/2$ ways to label the linear graph (the symmetry group is of order
$2$). So the contribution of the linear trees to $\monw$ decreases at most
geometrically with the size. As all trees give a 
nonnegative contribution, $\monw$ has its first singularity on the
real positive axis, and at a finite distance. 

In the situation when $\monw$ has a finite radius of convergence, we
conclude that there is a forest-like regime for the perturbed
Erd\"os-Renyi model that extends the forest-like regime of the pure
Erd\"os-Renyi model, and that it is described by the equations
eqs.(\ref{eq:mainu},\ref{eq:mainw},\ref{eq:mainuw}), at least in the
small $\alpha$ phase. We shall elaborate on this point in the sequel.

\subsection{Combinatorial remarks}

As we have seen before, the above formul\ae\ for the free energy rely
on crucial assumptions. What we would like to show in this subsection,
before embarking on a detailed discussion of analytic features of
these equations, is that the combinatorics embodied in
eq.(\ref{eq:mainuw}) is correct. Suppose that we forget about the
random graph model for a moment, and consider instead a random forest
model, $w$ being the generating function for random weighted trees.

Expand the function
$\monu(xe^z)$ in powers of $z$ : $\monu(xe^z)=\sum_l
\frac{z^l}{l!}\left(x\frac{d}{dx}\right)^l\monu(x)$. It is well-known
from quantum field theory that the formal expansion of
$$ \int dz
e^{\frac{1}{\hbar}\left(-\frac{z^2}{2}+\monu(xe^z)\right)}$$
is a weighted sum of all connected Feynmann graphs. The weight of a
Feynmann graph is computed as follows : each edge gives a factor
$\hbar$ (propagator), each vertex of degree $l$ gives a factor
$\hbar^{-1}\left(x\frac{d}{dx}\right)^l\monu(x)$  and finally one divides
by the order of the symmetry group of the graph. The logarithm is
given by the same sum, but restricted to connected graphs. For
connected graphs, the power of $\hbar$ is the number of loops minus
$1$, so the
dominant contribution in the small $\hbar$ limit restricts the sum to
connected loopless graphs, i.e. trees. On the other hand, the small
$\hbar$ limit is given by the saddle point approximation,
i.e. eq.(\ref{eq:mainuw}). So $\monw$ is a sum over all trees, each
vertex of degree $l$ giving a factor
$\left(x\frac{d}{dx}\right)^l\monu(x)$.

But $\monu(x)$ itself
is a tree generating function, so $\left(x\frac{d}{dx}\right)^l\monu(x)$
is the generating function for trees with $l$ marked vertices (a
vertex can be marked more than once). So
eq.(\ref{eq:mainuw}) means that to construct $\monw$, one takes arbitrary
trees, (call them naked trees) and then blows up every vertex of
degree $l$ into a new tree with $l$ marked vertices from which naked
edges emerge. Note that a naked vertex can be blown up in a trivial
tree, corresponding to the term $x$ in $\monu(x)=x+\cdots$. 

As we have emphasized before, if $T$ is a tree,
each term $t$ in the expansion of $e^{S_I(T)}$ in terms of matrix
elements of the adjacency matrix $A$ of $T$ defines a subgraph of $T$
i.e. a forest with the same vertex set as $T$, edge $\{i,j\}\in E(T)$
being present in the forest if and only if the term $t$ contains the
factor $A_{ij}$ or $A_{ji}$. But the connected components of the
forest being given, one reconstructs $\monw$ by connecting the
different components with appropriate edges.  This is exactly the
procedure described by eq.(\ref{eq:mainuw}) if $\monw$ is the
generating function for $e^{S_I(T)}$ and $\monu$ the one for
$\overline{e^{S_I(T)}}$.

\subsection{Effective model}

If we have the original model in mind, each $\monu_k$ is itself a highly
nontrivial kind of partition function. However, if we take  each
$\monu_k$ as an independent parameter, we can make a rather general
analysis. In fact, there is a simple model for which the $\monu_k$'s are
the fundamental microscopic parameters in the sense that they appear
directly in the definition of the weights. We call this model an
effective model for the following reasons. 

In quantum field theory, the term ``effective'' often means that one
renounces to deal with all observables of a system and only concentrates
on certain degrees of freedom, so that the other ones can be
averaged. For instance, to compute the long distance behaviour, one
first averages over the short distance fluctuations. We are going to do
something analogous here : we renounce to observe the local structure
of connected components and are only interested in the distribution of
their size. So instead of keeping track of the weight of each detailed
connected component, we can as well give all components of a given
size the same weight, namely the average weight given by the original
model for components of that size.

Now, to the precise definition.  Choose parameters $c_1,c_2,\cdots$
and define an effective weight $u^{(eff)}(H)=p^{E(H)}c_k$ for any
connected graph $H$ of size $k$, and assume multiplicativity, so that for
an arbitrary graph $u^{(eff)}(H)= p^{E(H)}\prod_k c_k^{n_k(H)}$ where
$n_k(H)$ is the number of components of size $k$ of $H$. If we trade
$c_k$ for $\lambda ^k c_k$, we multiply the weight $u^{(eff)}(H)$ by a
trivial factor $\lambda^{V(H)}$ so we shall assume that the $c_k$'s
are normalized by $c_1=1$ (the special case $c_1=0$ would need a
separate treatment). Define the corresponding effective weight
$w^{(eff)}(G)=q^{E(G)}\sum_{H,E(H)\subset E(G)}
u^{(eff)}(H)p^{-E(H)}$, where the sum is over all graphs on the same
vertex set as $G$ whose edge set is a subset of that of $G$. Note that
contrary to the weight $u^{(eff)}$, the weight $w^{(eff)}(G)$ does in
general depend on the detailed structure of the graph, and not only on
the sizes of connected components. Our interest however is in the
distribution of sizes of connected component of graphs of large size
$N \rightarrow \infty$ sampled using the weight $w^{(eff)}$. Following
the same steps as for the original model, we find that this
distribution can be obtained in the thermodynamic limit from tree
generating functions $\monu ^{(eff)}$ and $\monw ^{(eff)}$ satisfying
the very same coupled equations
eqs.(\ref{eq:mainu},\ref{eq:mainw},\ref{eq:mainuw}) as the original
$\monu$ and $\monw$.  The coefficients of $\monu ^{(eff)}$ are very
simple in terms of $c_1,c_2,\cdots$ because all components of the same
size have the same weight, and by Caley's theorem there are $k^{k-2}$
trees on $k$ vertices. Hence $\monu ^{(eff)}=\sum_{k \geq 1}
\frac{k^{k-2}}{k!}c_ky^k$. Hence if one sets $c_k=k^{2-k}\sum_{T \in
  \mathcal{T}_k} \overline{e^{S_I(T)}}$ where the sum is over trees of
size $k$, the effective model has the same component size distribution
as the original one.

For all these reasons, we shall remove in the sequel the superscript
$^{(eff)}$ from $\monu ^{(eff)}$ and $\monw ^{(eff)}$, even if we
sometimes keep the distinction between the weights $u$ and $w$ and the
effective weights $u^{(eff)}$ and $w^{(eff)}$. Accordingly, we
shall analyse eqs.(\ref{eq:mainu},\ref{eq:mainw},\ref{eq:mainuw}),
which involve only the component size distribution, without making
explicitly the distinction between the original model and the
effective model.

\subsection{Connected components and percolation}

We return to the finite $N$ arbitrary $p$ case to start the argument.
As $W_k$ is, modulo an overall multiplicative factor, the total
weight of connected graphs of size $k$, we infer from
eq.(\ref{eq:1expform}) that the mean number of connected components on
$k$ vertices is
\begin{eqnarray*}N_k & = & \frac{W_k}{Z_N} \frac{\partial Z_N}{\partial
 W_k}\\  & = & \frac{N!}{k!(N-k)!}W_k 
\frac{Z_{N-k}}{Z_N}(1-p)^{\frac{N(N-1)-(N-k)(N-k-1)}{2}}.
\end{eqnarray*}
Taking into account that when $pN=\alpha$, $p(N-k)=\alpha(1-k/N)$, we
find that in the dilute regime, for fixed $k$ and $N \rightarrow
\infty$, $\frac{Z_{N-k}}{Z_N}\sim e^{-k(F+\alpha \frac{\partial
    F}{\partial \alpha})}$ and
$(1-p)^{\frac{N(N-1)-(N-k)(N-k-1)}{2}}\sim e^{-k\alpha}$ so that
$$N_k/N \sim \frac{\monw_k}{k!} \alpha^{k-1}
e^{-k\left(\alpha+F+\alpha \frac{\partial F}{\partial \alpha}
  \right)}.$$
As expected, in this regime only trees contribute
thermodynamically to the finite components.

From these equations for the abundance of connected component of each
size, we can easily derive a percolation criterion. Indeed, by
construction, $\sum_k kN_k/N =1$, but what about the approximate sum 
$$\sum _k k\frac{\monw_k}{k!} \alpha^{k-1} e^{-k\left(\alpha+F+\alpha
    \frac{\partial F}{\partial \alpha} \right)} \; ?$$
For each fixed $k$
and $N\rightarrow \infty$, the $k^{th}$ term is a good approximation
to $ kN_k/N$, but there is problem of inversion of limits. Physically,
the approximate sum counts the fraction of points in components of
finite size, so it is $\leq 1$.

If we assume that $\monu$ is analytic at small $y$, then $F$ is
analytic and small at small $\alpha$ and $\monw$ is analytic and small
at small $x$. Moreover, $x^*$ is an increasing function of $\alpha$ at
small $\alpha$. From eq.(\ref{eq:mainw}), we infer that $F+\alpha
\frac{\partial F} {\partial \alpha}=-\alpha-\log \frac{x^*}{\alpha}$
or equivalently, $\alpha e^{-\alpha-F-\alpha \frac{\partial F}
  {\partial \alpha}}=x^*$. Then $\sum _k k\frac{\monw_k}{k!}
\alpha^{k-1} e^{-k\left(\alpha+F+\alpha \frac{\partial F} {\partial
      \alpha} \right)}= \frac{x^*}{\alpha}\monw '(x^*)=1$ for small
enough $\alpha$. However, it may happen that as a function of
$\alpha$, $x^*=\alpha e^{-\alpha-F-\alpha \frac{\partial F} {\partial
    \alpha}}$ is non monotonic. There may be a value $\alpha_c$ such
that $x^*$  increases in the interval $[0,\alpha_c]$ but then starts
to decrease, so that $x^*(\alpha) \leq x^*(\alpha_c)$ in some interval
strictly containing $[0,\alpha_c]$. One could build models where
$x^*(\alpha)$ has several oscillations, but in the sequel, we
concentrate on the first. For a given $\alpha$, denote by
$\bar{\alpha} \leq \alpha_c $ the small solution to the equation
$x^*(\alpha)=x^*(\bar{\alpha})$. Then we obtain the more general
result that finite components occupy a fraction
$\frac{x^*}{\alpha}\monw '(x^*)=\frac{\bar{\alpha}}{\alpha} \leq 1$ of
the sites in the system. If $\alpha > \alpha_c$, something else than
finite components, in fact on general grounds one single giant
component, occupies a fraction $1-\frac{\bar{\alpha}}{\alpha}$
vertices. Thus, the percolation criterion is that $\alpha
e^{-\alpha-F-\alpha \frac{\partial F} {\partial \alpha}}$ is maximum
at $\alpha=\alpha_c$. So the transition point is when 
\begin{equation}
  \label{eq:percoalpha}
\alpha+2\alpha
\frac{\partial F} {\partial \alpha}+\alpha^2 \frac{\partial^2 F}
{\partial \alpha^2}=1  
\end{equation}
The first two terms yield simply the true average connectivity $c =
\alpha +2\alpha \frac{\partial F}{\partial \alpha}$, it would be nice
to have a direct physical interpretation of the third term $\alpha^2
\frac{\partial^2 F} 
{\partial \alpha^2}$. This percolation criterion is expressed solely
in terms of the free energy as a function of $\alpha$. But it can also
be related to analytic properties of $\monw$. Indeed, the relevant
saddle point equation is $\alpha=x^*\monw'(x^*)$. As $\alpha$
approaches $\alpha_c$, $x^*$ reaches a maximum, so that the $x^*$
derivative of $x^*\monw'(x^*)$ has to get large, diverging at $\alpha
= \alpha_c$. If $\alpha_c$ is finite, this means that $\monw$ and
$\monw '$ are finite at $\alpha = \alpha_c$, but $\monw ''$ is
infinite. If the
coefficients of $\monw$ are non negative\footnote{This should be the
  case in statistical mechanics, and it is true by construction for
  our initial model as long as the parameters are real.}, this means that
$x^*(\alpha_c)$ is the radius of convergence of $\monw$. From
$x^*(\alpha_c)$, we recover $\alpha_c$ itself by the general saddle
point equation $\alpha=x^*\monw'(x^*)$.

Suppose now (we shall soon argue that this is true in many cases
including interesting ones) that even if $x^*(\alpha_c)$ is the radius
of convergence of $\monw$, the function $\monu$ is not singular at
$y^*(\alpha_c)$. Hence $\monu$ allows to compute the free energy $F$
and show that it is analytic in some interval strictly containing
$\alpha_c$. From that point of view, we observe that the saddle point
equations imply that $x^*=y^*e^{-\alpha}=y^*e^{-y^*\monu ' (y^*)}$
from which the percolation criterion, i.e. the determination of the
maximum of $x^*$ becomes
\begin{equation}
 \label{eq:percoy*}
 y^*\frac{\partial \alpha}{\partial y^*} = y^*\monu '
 (y^*)+{y^*}^2\monu '' (y^*)=1.  
\end{equation}
In the same spirit, the true average connectivity can be expressed as
$c = y^*\monu ' (y^*)+2-2\frac{\monu(y^*)}{y^*\monu ' (y^*)}$.

In general, if $\monu$ has nonegative coefficients and
eq.(\ref{eq:percoy*}) has a solution strictly within the disc of
convergence, one can go through the above argument in the reverse
order to prove the existence of a percolation transition with the
announced characteristics. This is the case for instance if $\monu$ is
an entire function with nonnegative coefficients, or more generally if
$\monu$ is function with nonnegative coefficients such that $\monu ''$
is unbounded when the argument approaches the radius of convergence. 
It is worth to observe that if the $y$ expansion of $\monu$ has
nonnegative coefficients, then the same is true of the $x$ expansion
of $\monw$. Indeed, from eq.(\ref{eq:mainuw}) and the corresponding
saddle point equation we infer that $x{\monw}'(x)=\hat{z}$. Hence as
functions of $x$, $\monw$ and $\hat{z}$ have the same singular points,
and
$$ {\monw}'(x)=e^{{x\monw}'(x)}{\monu}' (xe^{x{\monw}'(x)}).$$
Expand both sides of this identity to see that $w_1=1$ and that
$\monw_{k+1}-\monu_{k+1}$ is a polynomial in
$\monu_1=1,\monu_2,\cdots,\monu_k, \monw_1=1,\monw_2,\cdots,\monw_k$
with nonnegative coefficients. 

In the case of our original model, the situation is more tricky. We
know by construction that the $x$ expansion
of $\monw$ has nonnegative coefficients, but to ensure the existence
of a dilute regime, the same cannot be true in general of $\monu$.
In the
sequel, we shall see that in perturbation theory at any finite order,
we are in the following situation : the coefficients of $\monu$ may be 
negative, but nevertheless $\monu '(y)$ is analytic (in fact a
polynomial) and positive in a interval strictly containing $0$ and a
solution of eq.(\ref{eq:percoy*}). Then our previous arguments can be
made rigorous and there is a (perturbative) percolation transition
with the announced characteristics. We do not know if this argument
can be extended outside the realm of perturbation theory. The
numerical simulations are encouraging, but the behaviour of some
perturbative series is puzzling. Before discussing that, let
us consider three simple but significant examples. 

\section{Three easy examples}

\subsection{The case of the Erd\"os-Renyi model}

Let us recover the Erd\"os-Renyi model in
this framework. In that case, by construction, $U(y)=\monu(y)=y$ and
$y^*=\alpha$. Eq.(\ref{eq:mainu}) leads to $F=0$ for all values of
$\alpha$ (no surprise, for the Erd\"os-Renyi model the weights are
normalized as a probability distribution). Then eq.(\ref{eq:mainuw})
leads to $\hat{z}=xe^{\hat{z}}$, and from the Lagrange inversion formula, 
$$ \hat{z}=\sum_k \frac{k^{k-1}}{k!}x^k  \qquad \monw=\sum_k
\frac{k^{k-2}}{k!}x^k$$
which are the classical (rooted and non rooted) tree generating
functions (in fact, this gives a proof of Caley's formula for the
number of trees). Note that if we use naively eq.(\ref{eq:mainw}), we can deduce
that $F=0$ only for $\alpha \leq 1$. 

The number of connected components of size $k$ is 
$N_n\sim N
\frac{k^{k-2}}{k!}\alpha^{k-1}e^{-k\alpha}$, which is well-known to be
true for fixed $k$ and large $N$, for any value of $\alpha$. Notice
again that the use of $\monu$ plays a crucial role in our approach. Using
only $\monw$, we would get the component distribution only for $\alpha
\leq 1$. In fact, for the corresponding random
forest model (which is thermodynamically equivalent to the random
graph model for $\alpha \leq 1$) $\lim_{N\rightarrow \infty} N_n/N$ is
$\frac{k^{k-2}}{k!}\alpha^{k-1}e^{-k\alpha}$ for $\alpha \leq 1$ but
is nonanalytic at $\alpha=1$, which is the percolation
transition. 

The total number of points belonging to components of size $k$ is
$\sim N
\frac{k^{k-1}}{k!}\alpha^{k-1}e^{-k\alpha}$. 

For $\alpha \leq 1$, $\sum_k
\frac{k^{k-1}}{k!}\alpha^{k}e^{-k\alpha}=\alpha$,
but for $\alpha >  1$, $\sum_k
\frac{k^{k-1}}{k!}\alpha^{k}e^{-k\alpha}=\bar{\alpha}$, where
$\bar{\alpha}$ is the smallest solution to $\alpha
e^{-\alpha}=\bar{\alpha} e^{-\bar{\alpha}}$. The giant component
occupies $\sim N(1-\bar{\alpha}/\alpha)$ sites.

\subsection{The nested Erd\"os-Renyi model}

As another example, suppose that $u^{(eff)}(H)=p^{E(H)}$ for all
graphs, i.e that $c_k=1, k \geq 1$. Then
$w^{(eff)}(G)=q^{E(G)}\sum_{H,E(H)\subset E(G)} 1=(2q)^{E(H)}$.  Both
weights describe the Erd\"os-Renyi model, but with different values
for the probability of an edge. Going to the large $N$ finite
connectivity limit, we find $\monu=\sum _{k\geq 1}
\frac{k^{k-2}}{k!} y^k$, and from our previous analysis of the
Erd\"os-Renyi model, we find that $y {\monu}'(y)=\sum _{k\geq
  1} \frac{k^{k-1}}{k!} y^k$ is the Lambert function $L(y)$, the
solution of $L(y)e^{-L(y)}=y$ analytic close to $0$ and vanishing at
$0$.  Hence $\hat{z}=L(xe^{\hat{z}})$, so that
$\hat{z}e^{-\hat{z}}=xe^{\hat{z}}$. Hence $2\hat{z}=L(2x)$.  Moreover,
from $\monu(y)=L(y)- \frac{L(y)^2}{2}$ we find $2\monw
^{(eff)}(x)=L(2x)- \frac{L(2x)^2}{2}$. So we recover the doubling of
the edge probability when passing from the $u ^{(eff)}$ weight to the
$w ^{(eff)}$. The $u ^{(eff)}$ percolation transition is at $\alpha=1$
but the $w ^{(eff)}$ percolation transition occurs at $\alpha=1/2$.
Note that the equation $y^* {\monu}'(y^*)=L(y)=\alpha$ cannot
be solved for $\alpha \geq 1$, but that the free energy $F=\alpha/2$
and the true connectivity $c=2\alpha$ have an analytic continuation
for larger $\alpha$'s. That this analytic continuation is the true
value of $F$ cannot in principle be decided from our arguments (we
would have to do one more step of the same construction to view the
$u^{(eff)}$ weight itself as a composite weight).  But this does not
prevent us from finding and analysing correctly the $w ^{(eff)}$
transition, because it occurs strictly before the $u^{(eff)}$
transition.

\subsection{The matching model}

When $u_k=0$ for $k \geq 2$ we recover the Erd\"os-Renyi model, so let
us try the next degree of difficulty, when $u_k=0$ for $k \geq 3$ but
$u_2$ is a free parameter. Thus $w^{(eff)}(G)$ is the generating
function for a gas of disjoint egdes on $G$, that is, the generating
function for (all, non necessarily maximal) matchings on $G$. This is
a rather natural weight from the point of view of combinatorics. It is
plain that the detailed structure of $G$ is relevant, and not simply
the size of its connected components. On the other hand, the
$u^{(eff)}$ weight is nonzero only for a finite number of connected
graphs, so that the function $U^{(eff)}$ is simply $U^{(eff)}=z+qu_2
\frac{z^2}{2}$ and $Z_N=N!\oint \frac{dz}{z^{N+1}}e^{z+q u_2
  \frac{z^2}{2}}$. In such a simple case, the saddle point
approximation applies without subtleties, and we retrieve, in the
large $N$ finite connectivity limit, the expected equations. The
function $\monw (x)$ does not seem to be an elementary function. The
small $x$ and the perturbative small $u_2$ expansions are
straightforward but become quickly ugly. However from $\monu =y+u_2
\frac{y^2}{2}$, we can easily find the percolation criterion.
Parametrizing $u_2=\frac{1-y_c}{2y_c^2}$ (with $y_c \in ]0,1]$ for
positive $u_2$) and using eq.(\ref{eq:percoy*}), one finds that at the
percolation threshold :
$$y^*=y_c \qquad \alpha_c=\frac{1+y_c}{2} \qquad
c_{perc}=\frac{1+y_c}{2}+\frac{1-y_c}{1+y_c}.$$
So $\alpha_c$
decreases from $1$ to $1/2$ when $u_2$ grows, but the physical average
connectivity $c_{perc}$ increases from $1$ to $3/2$. The special case
$u_2=1$ is of special combinatorial significance, because the weight
$w^{(eff)}(G)$ counts the number of configurations of non adjacent
edges on $G$.  Then $y_c=1/2$, $\alpha_c=3/4$ and $c_{perc}=13/12$.
Consequently, $x_c=\frac{1}{2} e^{-3/4}$, from which we can derive a
result of direct combinatorial significance : 
$$\frac{1}{N!} \sum_{T\in {\mathcal T}_N} \# \ matchings \ of \ T \sim
C^{st} \frac{\left( 2e^{3/4}\right)^N}{N^{5/2}},$$
to be compared with
$\frac{1}{N!} \sum_{T\in {\mathcal T}_N} 1=\frac{N^{N-2}}{N!} \sim
\frac{1}{\sqrt{2\pi}} \frac{e^N}{N^{5/2}}$. Hence, if we put the
uniform probability law on labelled trees of size $N$, the average
number of matchings on a random tree of size $N$ behaves like $C^{st}
\left( \frac{16}{e} \right)^{N/4}$.

\section{Back to the original model}

\subsection{Finite orders in perturbation theory}

Remember that we established in section \ref{sec:expon} that
$$U(z)=z+\textstyle{\sum^c_H}
\frac{1}{s(H)}p^{e(H)} z^{v(H)} \prod _k t_k^{m_k(H)}s_k^{n_k(H)},$$
where $\sum^c_H$ is the sum over unlabelled connected admissible
graphs (we could equivalently reason in terms of normalized connected
sequences). Consider the coefficient of $\prod _k t_k^{m_k}s_k^{n_k}$ :
it is the sum over admissible graphs with $m_k$ solid $k$-cycles
and $n_k$ solid segments on $k+1$ vertices. There is only a
finite number of ways to join these fixed solid components with any
number of complete dashed graphs. So the coefficient of $\prod _k
t_k^{m_k}s_k^{n_k}$ is a polynomial in $p$ and $z$. A fortiori,
if we restrict to admissible graphs $H$ such that $v=l+1$, which are
the ones contributing to $\monu$, the sum is finite, and the coefficent
of $\prod _k t_k^{m_k}s_k^{n_k}$ in the perturbative expansion of
$\monu (y)$ is a polynomial in $y$. 

Note that $\monu (y)=y+O(y^2)$, so $y\monu ' (y)=y+O(y^2)$ and $y
\monu ' (y)+ y^2 \monu '' (y)=y+O(y^2)$, where the $O(y^2)$ vanish to
zeroth order in perturbation theory. Hence to any finite order in
perturbation theory, $y\monu '(y)$ is analytic and increasing up in a
large value of $y$, but $y\monu ' (y)+ y^2 \monu '' (y)=1$, the signal
of the percolation transition, occurs at a value of $y$ of order $1$.
Hence generically to any finite order in perturbation theory our
initial model exhibits a percolation transition described by our
previous results. In the following we shall make explicit perturbative
computations of the free energy, the percolation threshold, etc, for
the special case $t_k=0,s_k=2 \mu \delta_{k,3}$. To compare with the
prediction of the Molloy-Reed criterion, we need first to show how to
compute it in perturbation theory for our model.

\subsection{Moments of the degree distribution,
  Molloy-Reed's criterion}

By degree distribution of a given labelled graph $G$ on $N$ vertices
is meant the sequence $(n_0(G),n_1(G),\cdots,n_{N-1}(G))$ where
$n_i(G)$ is the number of vertices in $G$ with exactly $i$
neighbours. For fixed $N$, the
Molloy-Reed model concentrates on the set of all those labelled graphs
with a fixed degree distribution $(n_0,n_1,\cdots)$ and gives them uniform
probability, see \cite{molloy}. This represents a microcanonical point 
of view in the sense that the degree distribution is fixed and can not 
fluctuate. For a grand canonical presentation of the same idea, see
\cite{bb2}. If, for large $N$, $\left( n_0/N,n_1/N,\cdots \right)$
converges (in a sense made precise by Molloy and Reed) to a probability distribution $\left(
  f_0,f_1,\cdots \right)$, a limiting random graph model is obtained,
which depends only on $(f_0,f_1,\cdots)$ and not on the details of the 
approximating sequence $\left( n_0/N,n_1/N,\cdots \right)$. We now
recall the percolation criterion for the Molloy-Reed model with
arbitrary degree distribution.
\\
For a given graph $G$, define $k^{(q)}(G)$ as the following average over
vertices of $G$~:
$$k^{(q)}(G) \equiv \frac{1}{N} \sum_{i=1}^N l_i(G)^q.$$
For instance, when $q=1$, $Nk(G)=2E(G)$.
\\
The statistical average $\left< k^q \right> \equiv \frac{1}{\sum_G w(G)} \sum_G
w(G) k^{(q)}(G)$ is called the $q$-th moment of the degree
distribution. Note that in the Molloy-Reed model, all graphs have the
same degree distribution, so that $\left< k^q \right>=k^{(q)}(G)$ for all
$G$ in the relevant statistical ensemble.
\\
The Molloy-Reed percolation criterion states that the Molloy-Reed
random graph has a giant component if and only if the two first
moments of the degree distribution verify $\left< k^2-2k \right>
>0$. For the Erd\"{o}s-Renyi model, $\left< k
\right>=\alpha,\left< k^2 \right>=\alpha (\alpha+1)$, leading to the
percolation threshold $\alpha=1$.
\\
Our present purpose is to compute in perturbation theory the first
moments of the degree distribution for our model. In principle, it is possible to compute $\left<k^q \right>$ for any $q
\geq 1$. In the definition
$$\left< k^q \right> = \frac{1 } {N Z_N} \sum_{G \in {\mathcal G}_N} e^{S_I(G)} w_0(G)\sum_{i,j_1,\cdots,j_q} a_{ij_1} \cdots
a_{ij_q}$$
of the $q$-th moment, $e^{S_I(G)}\sum_{i,j_1,\cdots,j_q} a_{ij_1} \cdots
a_{ij_q}$ may be viewed as the $x$ derivative taken at $x=0$ of $\exp
\left( S_I^{(q)}(G,x) \right) \equiv \exp \left( S_I(G) + x \sum_{i,j_1,\cdots,j_q} a_{ij_1}
    \cdots a_{ij_q} \right)$. Seen as a new term of interaction, this
exponential is still multiplicative and permutation invariant. We thus
follow the steps which led us to eq.(\ref{eq:2expform}) (see
sec. \ref{reorg}) to prove that $\left< k^q \right>$ is the derivative
taken at $x=0$ of $\frac{(N-1)!}{Z_N} \oint \frac{dz}{z^{N+1}} e^{U^{(q)}(z)}=\frac{1}{N}\frac{Z_N^{(q)}}{Z_N},$ 
where $Z_N^{(q)}$ is the partition function of the model obtained from
the original model by replacing $S_I$ by  $S_I^{(q)}$. In the large $N$ limit, we proceed just as
in sec. \ref{maineq} to show that $Z_N^{(q)}=e^{NF_q+o(N)}$, $F_q$
being the new free energy : $F_q=-1-\log \frac
{y^\star_q}{\alpha} + \frac{1}{\alpha} \monu^{(q)}(y^\star_q)$. In this
expression, $\monu^{(q)}$ is the tree generating function for the new
model and $y_q^\star$ is the corresponding saddle point.
\\
We now take the derivative and put $x=0$ to yield
$$\left< k^q \right>=\frac{1}{\alpha} \sum_T \sum_{i,j_1,\cdots,j_q}\overline{e^{S_I(T)} a_{ij_1} \cdots
a_{ij_q}} \frac{y^{V(T)}}{V(T)!}$$

Just as we did in the original model, we can use normalized sequences
(or admissible graphs) to give a combinatorial interpretation of the
overlined term. A sequence $i_1j_1 \cdots i_nj_n$ is said to be normalized with respect
to $\{ m_k,n_k\},q$ if
\begin{itemize}
\item $n=q+\sum_k k \left( m_k+n_k \right)$, \\
\item 1 comes before 2, which comes before 3,... which comes before the
number $v$ of distinct elements among the sequence, \\
\item $i_1 \neq j_1, \cdots , i_n \neq j_n$, \\
\item it has a correct structure. That is, the sequence of the
  $2(n-q)$ first terms has a correct structure as regards $Tr$ and $\|
  \ \|$ and, moreover, $i_n=i_{n-1}=\cdots =i_{n-q+1}$.
\end{itemize}

We put ${\mathcal M}_{v,l,(m_k,n_k),q}$ for the number of such
sequences. Finally, the $q$-th moment of the degree distribution is
\begin{equation}
\left< k^q \right>=\frac{y^\star}{\alpha}+\frac{1}{\alpha}\sum_{v,\{m_\bullet,n_\bullet
  \}} \frac{\tilde{t}_\bullet^{m_\bullet}
\tilde{s}_\bullet^{n_\bullet} } {m_\bullet ! n_\bullet !} {\mathcal
  M}_{v,(m_k,n_k),q}^t {y^\star}^v
\label{momentq}
\end{equation}

In particular, the Molloy-Reed's criterion can, in principle, be
computed by means of this formula : it involves normalized
sequences (of type $(m_k,n_k)$) to which are concatenated
subsequences of 2 elements for $\left< k \right>$ or 3 elements for $\left< k^2
\right>$.

We now study a simple example in which all quantities mentioned above
can be explicitly (although perturbatively) computed.

\subsection{Perturbation theory : the example $t_k=0,s_k=2 \mu
  \delta_{k,3}$}
This is the simplest non-trivial case for which $s_k \neq 0$. The
weight of a graph $G$ is 
$$w(G)=p^{E(G)}(1-p)^{-E(G)} e^{\mu \sum_{ijkl}
  a_{ij}a_{jk}a_{kl}}.$$
According to our previous discussion, we assume that $\mu <0$. 
\begin{figure}
\begin{center}
\includegraphics[width=.95\textwidth]{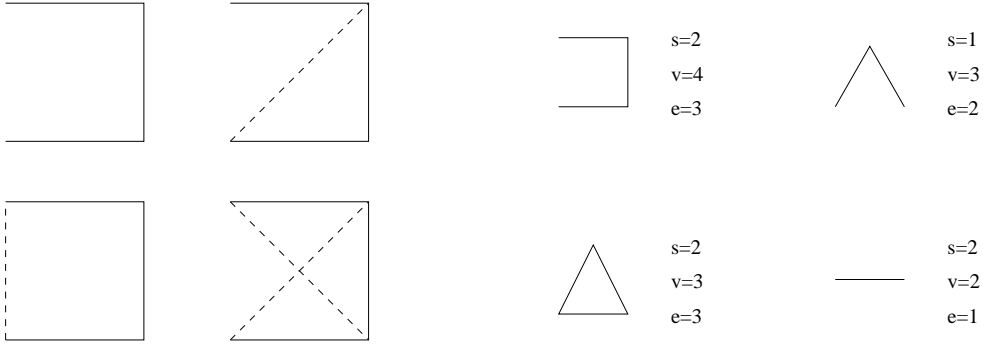}
\caption{Graphical representation for the coefficients of $s_3$.}
\label{coeffs3}
\end{center}
\end{figure}
To get the percolation criterion, we use the general theory exposed in
the preceding section. We set ${\mathcal M}^t_{n,m} \equiv
{\mathcal M}^c_{n,n-1,m}$. The definition of
$y^*$ is $y^* \monu ' (y^*) = \alpha$, i.e.
$$y^*+\sum_{n,m} n {y^*}^n \frac{\mu^m } {m!} {\mathcal
  M}^t_{n,m}=\alpha.$$

In fact, we were not able to find an expression
of ${\mathcal M}^t_{n,m}$ valid for all $n,m$, and we rely on a direct
enumeration, up to order 6, of the normalized sequences, see table
[\ref{tab:mtnp}] for the first five orders. At this moment, a fully
automated enumeration algorithm starting from scratch and
working in a reasonnable time would need too much memory. To
have some control over possible 
errors coming from human input, we have checked our results with two
independent algorithms. On a 2.5 Ghz processor, the computation of the
fifth  order takes
about 5 minutes, but the sixth order takes about 8 hours : the growth
in complexity is extremely rapid, at least factorial.

Up to third order in $\mu$, $y^\star$ takes the following form:
\begin{eqnarray*}
y^\star & = &
\alpha-2\alpha^2(\alpha+1)(2\alpha+1)\mu+2\alpha^2(-1-17\alpha-56\alpha^2-57\alpha^3-15\alpha^4 
\\
 & &
 +4\alpha^5)\mu^2-\frac{4}{3}\alpha^2(1+81\alpha+788\alpha^2+2485\alpha^3+3303\alpha^4+1808\alpha^5 \\
 & & +159\alpha^6-126\alpha^7+8\alpha^8)\mu^3+\cdots
\end{eqnarray*}

The percolation criterion states that there exists a giant connected
component if $y^*\monu '
 (y^*)+{y^*}^2\monu '' (y^*)>1$ and that, on the contrary, all connected components are of
finite size if $y^*\monu '
 (y^*)+{y^*}^2\monu '' (y^*)<1$. With $\monu (y)=y+\sum_{n,m}{\mathcal
  M}^t_{n,m} \frac{\mu^m}{m!} y^n$, the boundary between the percolating region and the non
percolating region is a curve in the $(\alpha,\mu)$ plane, of equation~:
\begin{equation}
\alpha-1+\sum_{n,m} n(n-1) {y^*}^n \frac{\mu^m } {m!} {\mathcal
  M}^t_{n,m}=0
\label{percolation}
\end{equation}

We can solve this equation for $\alpha$ as a perturbative series in $\mu$. Up to order 5, this yields
\begin{equation}
\alpha_{perc}=1-26 \mu+ 336 \mu^2-\frac{9500 } {3} \mu^3+\frac{49718} 
  {3} \mu^4-\frac{991328}{5} \mu^5-\frac{41436164}{15} \mu^6+\cdots
\label{alphaperc}
\end{equation}

\begin{table}
  \centering \tabcolsep 15pt
  \begin{tabular}{c|ccccc}
    $n\backslash m$ & 1 & 2 & 3 & 4 & 5 \\ \hline
    2 & 1 & 2 & 4 & 8 & 16 \\
    3 & 2 & 28 & 248 & 2032 & 16352 \\
    4 & 1 & 86 & 2236 & 44024 & 789616 \\
    5 &   & 108 & 7720 & 316784 & 10603040 \\
    6 &   &  66 & 14120 & 1152952 & 66713920 \\
    7 &   &  16 & 15424 & 2558624  & 248562304 \\
    8 &   &     & 10284 & 3781264 & 619455952 \\
    9 &   &     &  3888 & 3851664 & 1101864640 \\
    10 &  &     &   640 & 2698504 & 1444605680 \\
    11 &  &     &       & 1249712 & 1410932864 \\
    12 &  &     &       &  345600 & 1019814768 \\
    13 &  &     &       &   43264 & 531798240 \\
    14 &  &     &       &         & 189678720 \\
    15 &  &     &       &         & 41472000 \\
    16 &  &     &       &         & 4194304
  \end{tabular}
  \caption{\em ${\mathcal  M}^t_{n,m}$ for $m=1,\cdots,5$.}
  \label{tab:mtnp}
\end{table}
Putting $\alpha=\alpha_{perc}$ in formula (\ref{cavecF}) we find that, 
at the percolation threshold, the mean number of neighbours of a given
vertex is
\begin{equation}
c_{perc}=1-10\mu-50\mu^2-\frac{652}{3}\mu^3-\frac{19786}{3}\mu^4-\frac{3498268}{15}\mu^5-\frac{67025012}{9}\mu^6+\cdots
\label{cperc}
\end{equation}

In the preceding section, we saw how to infer the moments of the
degree distribution from enumeration of the appropriate normalized
sequences. Tables [\ref{tab:m1tnp}] and [\ref{tab:m2tnp}] show the
result of these enumerations for $\left< k \right>$ and $\left< k^2 \right>$.
\\
We compute  $\left< k \right>$ (either by means of formula
(\ref{cavecF}) or using the enumeration [\ref{tab:m1tnp}] together
with eq.(\ref{momentq})) and $\left< k^2 \right>$ as perturbative
series in $\mu$, and then solve the equation $\left< k^2-2k \right>=0$
in $\alpha$ to find
$$\alpha_{MR}=1-24 \mu+274 \mu^2-\frac{7324 } {3}
\mu^3+\frac{28708}{3} \mu^4-\frac{577988}{3} \mu^5+\cdots,$$
which does not coincide with $\alpha_{perc}$.

\begin{table}
  \centering \tabcolsep 15pt
  \begin{tabular}{c|ccccc}
    $n\backslash m$ & 1 & 2 & 3 & 4 & 5 \\ \hline
    2  & 2  &    4 &      8 & 16         & 32 \\
    3  & 12 &  120 &   1008 & 8160       & 65472 \\
    4  & 18 &  692 &  14952 & 276560     & 4836768 \\
    5  & 8  & 1600 &  80800 & 2902784    & 91337088 \\
    6  &    & 1844 & 225648 & 14935280   & 779078400 \\
    7  &    & 1080 & 375408 & 45982304   & 3849121728 \\
    8  &    &  256 & 392360 & 93526304   & 12533947744 \\
    9  &    &      & 255312 & 131789760  & 28896796992 \\
    10 &    &      &  95040 & 130610064  & 49053023200 \\
    11 &    &      &  15488 & 89956640   & 62460050560 \\
    12 &    &      &        & 41178240   & 59854882464 \\
    13 &    &      &        & 11291904   & 42704264192 \\
    14 &    &      &        & 1404928    & 22060944640 \\
    15 &    &      &        &            & 7812720000 \\
    16 &    &      &        &            & 1698693120 \\
    17 &    &      &        &            & 171051008 \\

  \end{tabular}
  \caption{\em Enumeration of the sequences appearing in
   $\left< k \right>$.}
  \label{tab:m1tnp}
\end{table}

\begin{table}
  \centering \tabcolsep 15pt
  \begin{tabular}{c|ccccc}
    $n\backslash m$ & 1 & 2 & 3 & 4 & 5 \\ \hline
    2  & 2  &    4 &       8 & 16        & 32 \\
    3  & 20 &  184 &    1520 & 12256     & 98240 \\
    4  & 44 & 1336 &   27440 & 500320    & 8725184 \\
    5  & 38 & 3812 &  171208 & 5937552   & 184842528 \\
    6  & 12 & 5676 &  546752 & 33681040  & 1713610432 \\
    7  &    & 4804 & 1060024 & 113992144 & 9088370528 \\
    8  &    & 2212 & 1341416 & 257520720 & 31755109024 \\
    9  &    &  432 & 1127280 & 410985696 & 79109699392 \\
    10 &    &      &  611232 & 474725904 & 146874463968 \\
    11 &    &      &  194480 & 397440176 & 207952308800 \\
    12 &    &      &   27648 & 236315376 & 226475616384 \\
    13 &    &      &         & 94941392  & 189140564736 \\
    14 &    &      &         & 23156224  & 119320803648 \\
    15 &    &      &         &  2592000  & 55138687200  \\
    16 &    &      &         &           & 17635164160  \\
    17 &    &      &         &           & 3491452928  \\
    18 &    &      &         &           & 322486272  \\

  \end{tabular}
  \caption{\em Enumeration of the sequences appearing in
  $\left< k^2 \right>$.}
  \label{tab:m2tnp}
\end{table}

\section{Discussion and perspectives}

In this paper, we have studied a class of perturbations of the
Erd\"{o}s-Renyi model which introduce correlations between the edges :
the weight of a graph depends on the abundance
of certain geometric features.

To solve this model, we have introduced an auxiliary model whose tree
generating function $\monu$ was expected to present better convergence 
properties than the original one $\monw$. The free energy $F$ in the
large $N$ limit has been determined and a percolation transition has
been established by means of an effective model : the percolation criterion is given by an
equation, either on $F$ or on $\monu$. We also have formul{\ae} for the
degree distributions.

On the basis of these general results we give explicit formul{\ae} for the above 
quantities in the particular case where all parameters but one
vanish. These perturbative results raise some crucial
questions. Indeed, we hope that the thermodynamical model makes sense
for $\mu<0$ but that $\mu>0$ has to be discarded because it gives too
much weight to strongly connected configurations and cannot be
treated like a diluted, tree-like, regime. In fact, up to sixth order, 
it is not so clear that the series for $\alpha_{perc}$ is actually convergent for negative 
$\mu$, because its general term increases very fast. However,
as suggested by the fifth and sixth terms, we hope that the following
terms may all be negative, the series hence being possibly
summable when $\mu<0$. This interpretation is supported by the form of the perturbative expansion (\ref{cperc}) of the physical connectivity
parameter $c_{perc}$, which seems much better behaved, with 
negative coefficients  for orders >0.

We also have computed the Molloy-Reed criterion, which does not give an appropriate
description of the percolation transition in this model. The $\mu$ expansion
of $\alpha_{MR}$ seems to present the same pathology as $\alpha_{perc}$. A possibility
is that this series is indeed divergent for negative $\mu$ : the
equation $\left< k^2-2k \right>$ may not admit any solution in
$\alpha$ as soon as $\mu<0$. Another possibility is that, just as for $\alpha_{perc}$ the series may stop to 
alternate at higher orders. Anyway, it would be
desirable to determine a class of models for which the Molloy-Reed criterion is valid, and we
believe that a minimal requirement may be a kind of locality. Indeed,
the Molloy-Reed criterion concentrates on the first two moments of the
degree distribution, which are local quantities in the sense that
$k^{(q)}(G)$ can be computed as soon as the immediate environment of each
vertex is known, independently of how the vertices are connected to
each other. Even in the simple model that we used to illustrate
perturbation theory, this information is not sufficient to
compute the weight of a graph : one must also know the immediate
environment of the first neighbours of each vertex.
\\
Finally, we also believe that a more thorough understanding of degree
correlations induced by attacks deserves a systematic treatment.

\end{document}